\documentclass[conference]{IEEEtran}
\usepackage{cite}
\usepackage[cmex10]{amsmath}
\usepackage{amssymb,bm,comment}

\newtheorem{theorem}{Theorem}
\newtheorem{proposition}{Proposition}
\newtheorem{lemma}{Lemma}
\newtheorem{definition}{Definition}

\begin{document}
\title{Refined Rate of Channel Polarization}
\author{\IEEEauthorblockN{Toshiyuki Tanaka and Ryuhei Mori}
\IEEEauthorblockA{Graduate School of Informatics, Kyoto University, Kyoto-shi, Kyoto, 606-8501 Japan.\\
 e-mail: tt@i.kyoto-u.ac.jp, rmori@sys.i.kyoto-u.ac.jp}}
\maketitle

\begin{abstract}
A rate-dependent upper bound of the best achievable block error probability 
of polar codes with successive-cancellation decoding is derived.  
\end{abstract}

\maketitle

\section{Introduction}
Channel polarization~\cite{Arikan2009} is a method 
which allows us to construct a family of error-correcting codes, called polar codes.  
Polar codes have been attracting theoretical interest because 
they are capacity achieving for binary-input symmetric memoryless channels (B-SMCs), 
which are also achieving symmetric capacity for general binary-input memoryless channels (B-MCs), 
whereas computational complexity of encoding and decoding is polynomial in the block length.  
Soon after the first proposal~\cite{Arikan2009}, one can find in the literature 
a number of contributions regarding channel polarization and polar codes~\cite{%
Arikan2008,MoriTanaka2009b,ArikanTelatar2009,KoradaSasoglu2009,KoradaSasogluUrbanke2009,
HussamiKoradaUrbanke2009,MoriTanaka2009a,KoradaUrbanke2009pre,HassaniKoradaUrbanke2009pre}.  

Of particular theoretical interest is 
analysis of how fast the best achievable block error probability $P_e$ of polar codes 
decays toward zero as the block length $N$ tends to infinity.  
Ar\i kan~\cite{Arikan2009} has shown that $P_e$ tends to zero as $N\to\infty$ 
whenever the code rate $R$ is less than the symmetric capacity 
of the underlying channel.  
The upper bound he obtained is proportional to a negative power of $N$, 
which means that guaranteed speed of the convergence to zero is very slow.  
His result has subsequently been improved by Ar\i kan and Telatar~\cite{ArikanTelatar2009}, 
who have obtained a much tighter upper bound, which scales as exponential in $-N^\beta$ for $\beta<1/2$.  
Both of these bounds, however, do not depend on the code rate $R$.  
A rate-dependent bound is more desirable, 
since one naturally expects a smaller error probability from a smaller code rate, 
which might in turn suggest that the rate-independent bounds are not tight.  

In this paper, we present an analysis of the rate 
of channel polarization.  
The argument basically follows that of Ar\i kan and Telatar~\cite{ArikanTelatar2009}, 
but extends it to obtain rate-dependent bounds of the best achievable 
error probability.  

\section{Problem}
\label{sec:prob}
Let $W:\ {\cal X}\mapsto{\cal Y}$ be an arbitrary binary-input 
memoryless channel (B-MC) with input alphabet ${\cal X}=\{0,\,1\}$, 
output alphabet ${\cal Y}$, and channel transition probabilities 
$\{W(y|x):\ x\in{\cal X},\;y\in{\cal Y}\}$.  
Let $I(W)$ be the symmetric capacity of $W$, which is defined as 
the mutual information between the input and output of $W$ 
when the input is uniformly distributed over ${\cal X}$.  
It is an upper bound of achievable rates over $W$ 
with codes that use input symbols with equal frequency.  
Let the Bhattacharyya parameter $Z(W)$ of the channel $W$ be 
defined as 
\[
Z(W)=\sum_{y\in{\cal Y}}\sqrt{W(y|0)W(y|1)}.
\]
It is an upper bound of the maximum-likelihood estimation error 
for a single channel usage.  

Polar codes are constructed on the basis of recursive application 
of channel combining and splitting operation.  
In this operation, two independent copies of a channel $W$ is combined and then 
split to generate two different channels $W^-:\;{\cal X}\to{\cal Y}^2$ 
and $W^+:\;{\cal X}\to{\cal Y}^2\times{\cal X}$.  
The operation, in its most basic form, is defined as 
\begin{align}
W^-(y_1,\,y_2|x_1)&=\sum_{x_2\in{\cal X}}\frac{1}{2}W(y_1|(\bm{x}F)_1)W(y_2|(\bm{x}F)_2),
\nonumber\\
W^+(y_1,\,y_2,\,x_1|x_2)&=\frac{1}{2}W(y_1|(\bm{x}F)_1)W(y_2|(\bm{x}F)_2),
\end{align}
with 
\begin{equation}
\label{eq:mtxF}
F=\left(
\begin{array}{cc}
1&0\\1&1
\end{array}
\right),\quad
\bm{x}=(x_1,\,x_2).
\end{equation}
It has been shown~\cite{Arikan2009} that 
\begin{align}
\label{eq:recursion}
Z(W^+)&=Z(W)^2,
\nonumber\\
Z(W)\le Z(W^-)&\le2Z(W)-Z(W)^2.
\end{align}
In constructing polar codes, we recursively generate channels 
with the channel combining and splitting operation, 
starting with the given channel $W$, as 
\begin{align}
W
&\to\{W^-,\,W^+\}
\to\{W^{--},\,W^{-+},\,W^{+-},\,W^{++}\}
\nonumber\\
&\to\{W^{---},\,W^{--+},\,W^{-+-},\,W^{-++},
\nonumber\\
&\hphantom{\to\{}W^{+--},\,W^{+-+},\,W^{++-},\,W^{+++}\}
\to\cdots,
\end{align}
where we have adopted the shorthand notation $W^{--}=(W^-)^-$, etc.  

Following Ar\i kan~\cite{Arikan2009}, 
this process of recursive generation of channels can be 
dealt with by introducing a channel-valued stochastic process, 
defined as follows.  
Let $\{B_1,\,B_2,\,\ldots\}$ be a sequence of independent and identically 
distributed (i.i.d.) Bernoulli random variables with $P(B_1=0)=P(B_1=1)=1/2$.  
Given a channel $W$, we define a sequence of channel-valued random variables 
$\{W_0,\,W_1,\,\ldots\}$ as 
\begin{equation}
W_0=W,\quad W_{n+1}=\left\{
\begin{array}{ll}
W_n^- & \hbox{if $B_{n+1}=1$}\\
W_n^+ & \hbox{if $B_{n+1}=0$}
\end{array}
\right..
\end{equation}
We also define a real-valued random process $\{Z_0,\,Z_1,\,\ldots\}$ 
via $Z_n=Z(W_n)$.  

Conceptually, a polar code is constructed by picking up 
channels with good quality, among $N=2^n$ realizations of $W_n$.  
We use these selected channels for transmitting data, while some 
predetermined values are transmitted over the remaining unselected channels.  
Thus, the rate of the resulting polar code is $R$ if we pick up $NR$ channels.  
We are interested in performance of polar codes under successive cancellation (SC) 
decoding, which is defined in~\cite{Arikan2009}.  
Let $P_e(N,\,R)$ be the best achievable block error probability 
of polar codes of block length $N$ and rate $R$ 
under successive cancellation decoding.  
Since the Bhattacharyya parameter $Z(W)$ serves as an upper bound 
of bit error probability in each step of successive cancellation decoding, 
an inequality of the form 
\begin{equation}
\label{eq:generic}
P(Z_n\le\gamma)\ge R
\end{equation}
implies $P_e(N,\,R)\le NR\gamma$ via union bound.  

It has been proved~\cite{Arikan2009} that, for any $R<I(W)$, 
there exists a polar code with block length $N=2^n$, 
whose block error probability $P_e(N,\,R)$ is arbitrarily close to 0.  
The proof is based on showing the condition~\eqref{eq:generic} 
to hold for $\gamma\in o(N^{-1})$.  

\section{Main Result}
\label{sec:mainres}
The main contribution of this paper is to prove the following theorem, 
which improves the results in~\cite{Arikan2009,ArikanTelatar2009}, 
giving a rate-dependent upper bound of the block error probability.  

\begin{theorem}
\label{th:Pe}
Let $W$ be any B-MC with $I(W)>0$.  
Let $R\in(0,I(W))$ be fixed.  
Then, for $N=2^n$, $n\in\mathbb{N}$, 
the best achievable block error probability $P_e(N,\,R)$ satisfies, 
\begin{equation}
P_e(N=2^n,\,R)=o\left(2^{-2^{(n+t\sqrt{n})/2}}\right),
\end{equation}
for any $t$ satisfying $t<Q^{-1}(R/I(W))$, 
where $Q(x)=\int_x^\infty e^{-u^2/2}\,du/\sqrt{2\pi}$.  
\end{theorem}

\section{Proof}
\label{sec:proof}
\subsection{Outline}
\label{sec:outline}
The proof basically follows that of Ar\i kan and Telatar~\cite{ArikanTelatar2009} 
but extends it in several respects.  
It consists of three stages, 
which we call polarization, concentration, and bootstrapping, respectively.  
In the first stage, it will be argued that 
realizations of $Z_n$ are in $(0,\,\zeta]$ for some $\zeta>0$
with probability arbitrarily close to $I(W)$ as $n$ becomes large.  
This corresponds to the fundamental result of channel polarization~\cite{Arikan2009}.  
In the second stage, concentration will be argued, that is, 
again with probability arbitrarily close to $I(W)$ as $n$ gets large, 
realizations of $Z_n$ are in $(0,\,f_n]$ for 
some $f_n$ approaching zero exponentially in $n$.  
In the last stage, we will argue that, once $Z_m$ for some $m$ 
enters the interval $(0,\,f_n]$, the sequence $Z_{m+1},\,\ldots,\,Z_n$ is 
rapidly decreasing with overwhelming probability, 
which is a refinement of the ``bootstrapping argument'' of~\cite{ArikanTelatar2009}.  
The last stage is further divided into two substages, 
the rate-independent bootstrapping stage and the rate-dependent bootstrapping stage, 
the latter of which is crucial in order to see 
dependence on the code rate.  

\subsection{Preliminaries}
\label{sec:pre}
For $m,\,n\in\mathbb{N}$ with $m<n$, define 
\begin{equation}
\label{eq:def-S}
S_{m,\,n}=\sum_{i=m+1}^nB_i,
\end{equation}
which follows a binomial distribution, since 
it is a sum of i.i.d.\ Bernoulli random variables.  
\begin{definition}
For a fixed $\gamma\in[0,\,1]$, 
let $\mathcal{G}_{m,\,n}(\gamma)$ be the event defined by 
\[
\mathcal{G}_{m,\,n}(\gamma)=\{S_{m,\,n}\ge\gamma(n-m)\}.
\]
\end{definition}
From the law of large numbers, 
\begin{equation}
\label{eq:G-lln}
\lim_{n-m\to\infty}P(\mathcal{G}_{m,\,n}(\gamma))=1
\end{equation}
holds if $\gamma< 1/2$.  

\subsection{Random Process}
\label{sec:rp}
We now consider the random process $X_n\in[0,1]$ satisfying 
the following properties.  
\begin{enumerate}
\item $X_n$ converges to a random variable $X_\infty$ almost surely.
\item Conditional on $X_n$, if $X_n\not=0,\,1$, 
\[
X_{n+1}\left\{
\begin{array}{ll}
\in[X_n,\,qX_n] & \hbox{if $B_{n+1}=1$}\\
=X_n^2 & \hbox{if $B_{n+1}=0$}
\end{array}\right.
\]
for a constant $q\ge1$, and $X_{n+1}=X_n$ with probability 1 for $X_n=0$ or 1.  
\end{enumerate}
Equation~\eqref{eq:recursion} implies that the random process $Z_n$ satisfies the above properties 
with $q=2$.  
It should be noted that the properties 1 and 2 imply $P(X_\infty \in \{0,1\}) = 1$.  

\begin{definition}
\label{def:polarization}
For $\zeta\in(0,\,1)$ and $n\in\mathbb{N}$, define an event ${\cal T}_n(\zeta)$ with 
\[
{\cal T}_n(\zeta)=\{X_i\le\zeta;\,{}^\forall i\ge n\}.
\]
\end{definition}

The following lemma is an immediate consequence of the above definition. 

\begin{lemma}
\label{lemma:polarization}
For any fixed $\zeta\in(0,\,1)$, 
\[
\lim_{n\to\infty}P({\cal T}_n(\zeta))=P(X_\infty = 0).
\]
\end{lemma}

\subsection{Concentration}
For large enough $n$, 
one can expect that $X_n$ is exponentially small in $n$ 
with probability arbitrarily close to $P(X_\infty=0)$.  
In other words, a $P(X_\infty=0)$-fraction of realizations of $X_n$ 
``concentrates'' toward zero.   
To formalize the above statement, 
we introduce the following definition.  
\begin{definition}
Let $\rho\in(0,\,1)$ and $\beta\in(0,\,1/2)$.  
The events $\mathcal{C}_n(\rho)$ and $\mathcal{D}_n(\beta)$ are defined as 
\begin{align}
\label{eq:def-C}
\mathcal{C}_n(\rho)&=\{X_n\le\rho^n\},\\
\label{eq:def-D}
\mathcal{D}_n(\beta)&=\{X_n\le 2^{-2^{n\beta}}\},
\end{align}
respectively.  
\end{definition}
We will first prove that the event $\mathcal{C}_n$ 
has a probability 
arbitrarily close to $P(X_\infty=0)$ as $n$ tends to infinity, 
on the basis of which we will next prove that 
the event $\mathcal{D}_n(\beta)$ has a probability 
arbitrarily close to $P(X_\infty=0)$ as $n\to\infty$.  

The result for the event $\mathcal{C}_n$ is proved in the following proposition, 
on the basis of which the result for the event $\mathcal{D}_n$ is proved 
in the bootstrapping stage.  

\begin{proposition}
\label{prop:concentration}
For an arbitrary fixed $\rho\in(0,\,1)$, 
let $\mathcal{C}_n(\rho)$ be the event defined as~\eqref{eq:def-C}.  
Then,  
\[
\lim_{n\to\infty}P\left(\mathcal{C}_n(\rho)\right)= P(X_\infty=0).
\]
\end{proposition}
The proof is essentially the same as that for Theorem 2 in~\cite{Arikan2009}, 
and is omitted due to space limitations.  

\subsection{Bootstrapping: Rate-Independent Stage}
\label{sec:b-ri}
For some $m\ll n$, once a realization of $X_m$ becomes small enough, 
one can assure, with probability very close to 1, 
that samples conditionally generated on the realization of $X_m$ 
will converge to zero exponentially fast.  
This is the basic idea leading to the ``bootstrapping argument'' 
of~\cite{ArikanTelatar2009}.  
We basically follow the same idea.  

The proof regarding the bootstrapping stage is based on 
a consideration of properties of a process $\{L_i\}$, 
defined on the basis of $\{X_i\}$ as 
\begin{align}
\label{eq:def-Ln1}
L_i&=\log_2X_i,\quad i=0,\,\ldots,\,m,\\
\label{eq:def-Ln2}
L_{i+1}&=\left\{
\begin{array}{ll}
2L_i & \hbox{if $B_{i+1}=1$}\\
L_i+\log_2q & \hbox{if $B_{i+1}=0$}
\end{array}
\right.,\quad i\ge m
\end{align}
for a fixed $m$.  
The inequality $X_i\le 2^{L_i}$ holds on the sample-path basis 
for all $i\ge0$.  

If we fix $L_m$ and $S_{m,\,n}$, the largest value of $L_n$ 
is achieved by the sequence $\{B_{m+1},\,\ldots,\,B_n\}$ 
of $(n-m-S_{m,\,n})$ consecutive 0s followed by 
$S_{m,\,n}$ consecutive 1s.  
One therefore obtains 
\begin{equation}
\label{eq:ub-Ln}
L_n \le 2^{S_{m,\,n}}\left[L_m + (n - m - S_{m,\,n})\log_2q\right].
\end{equation}

\begin{lemma}
\label{lemma:Ln}
Fix $\gamma\in[0,\,1]$ and $\varepsilon>0$, 
and let $\rho=\rho(\gamma)$ be such that 
$\log_2\rho=-(1-\gamma)(n-m)\log_2q/m-\varepsilon$ holds.  
Then, conditional on ${\cal C}_m\bigl(\rho(\gamma)\bigr)\cap{\cal G}_{m,\,n}(\gamma)$, 
the inequality 
\[
L_n\le -2^{\gamma(n-m)}\varepsilon m
\]
holds.  
\end{lemma}
\begin{IEEEproof}
Conditional on ${\cal C}_m(\rho)\cap{\cal G}_{m,\,n}(\gamma)$, one has, from~\eqref{eq:ub-Ln}, 
the inequality 
\[
L_n\le 2^{S_{m,\,n}}\left[m\log_2\rho+(1-\gamma)(n-m)\log_2q\right].
\]
Letting $\rho=\rho(\gamma)$ completes the proof.  
\end{IEEEproof}

\begin{proposition}
\label{prop:boot1}
For an arbitrary fixed $\beta\in(0,\,1/2)$, 
let $\mathcal{D}_n(\beta)$ be the event defined in~\eqref{eq:def-D}.  
Then, 
\[
\lim_{n\to\infty} P\left(\mathcal{D}_n(\beta)\right) = P(X_\infty=0).
\]
\end{proposition}
\begin{IEEEproof}
Since $\beta\in(0,\,1/2)$, 
there exists $(\gamma,\,\alpha)\in(0,\,1/2)\times(0,\,1)$ 
satisfying the condition $\gamma(1-\alpha)=\beta$  
(e.g., letting $\gamma=(1+2\beta)/4$ and $\alpha=(1-2\beta)/(1+2\beta)$ 
satisfies the condition).  
We take $m=\alpha n$ in Lemma~\ref{lemma:Ln}, 
and let $\{L_i^{(1)}\}$ denote the process defined by~\eqref{eq:def-Ln1} and \eqref{eq:def-Ln2} with $m=\alpha n$.  
Then, for any $\varepsilon>0$, 
one obtains by applying Lemma~\ref{lemma:Ln} that, 
conditional on the event ${\cal C}_{\alpha n}\bigl(\rho(\gamma)\bigr)\cap{\cal G}_{\alpha n,\,n}(\gamma)$ 
with $\rho(\gamma)$ defined in Lemma~\ref{lemma:Ln}, 
the inequality 
\[
L_n^{(1)}\le-2^{\gamma(1-\alpha)n}\varepsilon\alpha n
\]
holds, which in turn implies 
\begin{equation}
\label{eq:incl-rel1}
\left\{X_n\le 2^{-2^{\gamma(1-\alpha)n}\varepsilon\alpha n}\right\}
\supset{\cal C}_{\alpha n}\bigl(\rho(\gamma)\bigr)\cap{\cal G}_{\alpha n,\,n}(\gamma).
\end{equation}

For any $n\ge(\varepsilon\alpha)^{-1}$, 
$\beta n\le \gamma(1-\alpha)n+\log_2\varepsilon\alpha n$ holds, 
so that one obtains 
\begin{equation}
\label{eq:incl-rel2}
\mathcal{D}_n(\beta)\supset\left\{X_n\le 2^{-2^{\gamma(1-\alpha)n}\varepsilon\alpha n}\right\}.
\end{equation}
From~\eqref{eq:incl-rel1} and \eqref{eq:incl-rel2}, 
as well as the independence of ${\cal C}_{\alpha n}\bigl(\rho(\gamma)\bigr)$ 
and ${\cal G}_{\alpha n,\,n}(\gamma)$, 
one consequently has 
\begin{equation}
P(\mathcal{D}_n(\beta))\ge P(\mathcal{G}_{\alpha n,\,n}(\gamma))
P(\mathcal{C}_{\alpha n}(\rho(\gamma))).
\end{equation}

Hence, using~\eqref{eq:G-lln} and Proposition~\ref{prop:concentration}, 
\begin{multline}
\lim_{n\to\infty}P(\mathcal{D}_n(\beta))
\ge\lim_{n\to\infty}P\bigl(\mathcal{G}_{\alpha n,\,n}(\gamma)\bigr)
P\bigl(\mathcal{C}_{\alpha n}\bigl(\rho(\gamma)\bigr)\bigr)\\
=P(X_\infty=0).
\end{multline}
\end{IEEEproof}

\subsection{Bootstrapping: Rate-Dependent Stage}
\label{sec:b-rd}
So far, our treatment of the random variable $S_{m,\,n}$ is restricted 
to that within regimes of the law of large numbers.  
In order to obtain a rate-dependent bound, 
we have to go further and treat $S_{m,\,n}$ within regimes of the central limit theorem.  
\begin{definition}
\label{def:H}
For $t\in\mathbb{R}$ and for a function $f(n)=o(\sqrt{n})$, 
the event ${\cal H}_{m,\,n}(t)$ is defined as 
\begin{equation}
\label{eq:def-H}
\mathcal{H}_{m,\,n}(t)=\left\{S_{m,\,n} \ge \frac12 (n-m + t\sqrt{n-m}) + f(n-m)\right\}.
\end{equation}
\end{definition}
Noting that the random variable $S_{m,\,n}$ is a sum of 
$(n-m)$ i.i.d.\ Bernoulli random variables, 
and that the mean and the variance are $(n-m)/2$ and $(n-m)/4$, respectively, 
the following lemma is a direct consequence of the central limit theorem.  
\begin{lemma}Let $m<n$.  
Then, for any $t\in\mathbb{R}$, 
\[
\lim_{n-m\to\infty}P(\mathcal{H}_{m,\,n}(t))=Q(t).
\]
\end{lemma}

\begin{proposition}
\label{prop:direct}
For an arbitrary function $f(n)=o(\sqrt{n})$.
\[
\liminf_{n\to\infty} P\left(X_n\le 2^{-2^{(n+t\sqrt{n})/2+f(n)}}\right) \ge Q(t)P(X_\infty=0).
\]
\end{proposition}
\begin{IEEEproof}
For a fixed $\beta\in(0,\,1/2)$, 
we take $m=\frac1\beta\log_2 n$ in Lemma~\ref{lemma:Ln}, 
and let $\{L_i^{(2)}\}$ denote the process defined by~\eqref{eq:def-Ln1} and \eqref{eq:def-Ln2} 
with this choice of $m$.  

Conditional on the event $\mathcal{D}_m(\beta)$, one obtains, from~\eqref{eq:ub-Ln}, the inequality 
\begin{equation}
L_n^{(2)} \le 2^{S_{m,\,n}}\left[-2^{\beta m}
+ (n - m - S_{m,\,n})\log_2q\right].
\end{equation}
Let $\mathcal{H}_{m,\,n}(t)$ be the event defined in Definition~\ref{def:H} 
for a fixed $t\in\mathbb{R}$ and for an arbitrarily chosen function $f(n)=o(\sqrt{n})$.
Conditional on $\mathcal{D}_m(\beta)\cap\mathcal{H}_{m,\,n}(t)$, $L_n^{(2)}$ is bounded 
from above as 
\begin{multline}
L_n^{(2)} \le 2^{\frac12(n-m)+\frac12t\sqrt{n-m}+f(n-m)}\\
 \times \left[-2^{\beta m} + \left(\frac{n-m-t\sqrt{n-m}}{2}-f(n-m)\right)\log_2q\right]\nonumber
\end{multline}
which implies that there exists $n_0$ such that for all $n\ge n_0$, 
the condition 
\begin{equation}
\left\{X_n \le 2^{-2^{\frac12(n-m)+\frac12t\sqrt{n-m}+f(n-m)}}\right\}
\supset \mathcal{D}_m(\beta)\cap\mathcal{H}_{m,\,n}(t)
\end{equation}
is satisfied.  
From this observation, as well as the independence 
of $\mathcal{D}_m(\beta)$ and $\mathcal{H}_{m,\,n}(t)$, 
one has 
\begin{multline}
P\left(X_n \le 2^{-2^{\frac12(n-m)+\frac12t\sqrt{n-m}+f(n-m)}}\right)\\
\ge P(\mathcal{D}_m(\beta))P(\mathcal{H}_{m,\,n}(t)).
\end{multline}
Thus, 
\begin{multline}
\liminf_{n\to\infty}
P\left(X_n \le 2^{-2^{\frac12(n-m)+\frac12t\sqrt{n-m}+f(n-m)}}\right)\\
\ge \lim_{n\to\infty} P(\mathcal{D}_m(\beta))P(\mathcal{H}_{m,\,n}(t))
= P(X_\infty=0)Q(t).
\end{multline}
Since $m=o(\sqrt{n})$, one can safely absorb possible effects of $m$ into 
the function $f$.  
This completes the proof.  
\end{IEEEproof}

\subsection{Converse}
In this subsection, we discuss the converse, 
in which probabilities that $X_n$ takes small values 
are bounded from above.  
\begin{proposition}
\label{prop:converse}
For an arbitrary function $f(n)=o(\sqrt{n})$ 
\[
\mathop{\lim\sup}_{n\to\infty} P\left(X_n\le 2^{-2^{(n+t\sqrt{n})/2+f(n)}}\right)\le Q(t)P(X_\infty=0)
\]
\end{proposition}
\begin{IEEEproof}
Fix a process $\{X_n\}$.  
Let $\{\check{X}_n\}$ be the random process defined as 
\begin{align}
\check{X}_i &= X_i,&\text{for } i = 0,\cdots,m\\
\check{X}_i &= \begin{cases}
\check{X}_{i-1}^2, & \text{if } B_i = 1\\
\check{X}_{i-1}, & \text{if } B_i = 0
\end{cases},& \text{for } i>m
\end{align}
The inequality $X_i\ge\check{X}_i$ holds on the sample-path basis for all $i\ge0$, 
which implies 
\[
P(X_n\le a)\le P(\check{X}_n\le a),
\]
for any $a$.  
One also has 
\[
\log_2\log_2(1/\check{X}_{m+k})=S_m^{m+k}+\log_2\log_2(1/X_m)
\]
The central limit theorem dictates that $\frac{2}{\sqrt{k}}(S-k/2)$ asymptotically follows 
the standard Gaussian distribution, so that, 
for any fixed $m$ and for an arbitrary function $f(k)=o(\sqrt{k})$, 
one has 
\begin{multline}
P\left(\check{X}_{m+k}\le 2^{-2^{(k+t\sqrt{k})/2+f(k)}}\Bigm|X_m\right)
\\
=P\left(\log_2\log_2(1/\check{X}_{m+k})\le \frac{k}{2}+\frac{t\sqrt{k}}{2}+f(k)
\Biggm|X_m\right)\\
= Q(t) + o(1)\label{eq:clt}
\end{multline}
as $k\to\infty$.
For any fixed $\delta\in(0,1)$, and $m\ge 0$
\begin{multline}
\label{eq:conv-ub}
\limsup_{n\to\infty} P\left(X_{n} \le 2^{-2^{(n + t\sqrt{n})/2+f(n)}}\right)\\
\le
\limsup_{k\to\infty} P\left(X_{m+k} \le 2^{-2^{(m+k + t\sqrt{m+k})/2+f(m+k)}}\right)\\
\le \mathop{\lim\sup}_{k\to\infty}\biggl\{
P\left(\check{X}_{m+k} \le 2^{-2^{(m+k + t\sqrt{m+k})/2+f(m+k)}}\biggm| X_m\le\delta\right)\\
\times P(X_m\le\delta)
+ P\left(X_{m+k} \le \frac{\delta}{2},~X_m>\delta\right)\biggr\}.
\end{multline}
From Fatou's lemma,
\begin{equation}
\label{eq:ub-sq1d2}
\limsup_{k\to\infty}P\left(X_{m+k} \le \frac{\delta}{2},~X_m>\delta\right)
\le P\left(X_\infty \le \frac{\delta}{2},~X_m>\delta\right).
\end{equation}
On the basis of~\eqref{eq:clt}, \eqref{eq:conv-ub}, and \eqref{eq:ub-sq1d2}, one obtains 
\begin{multline}
\limsup_{n\to\infty} P\left(X_{n} \le 2^{-2^{(n + t\sqrt{n})/2+f(n)}}\right)\\
\le Q(t)P(X_m\le\delta) + P\left(X_\infty \le \frac{\delta}{2},~X_m>\delta\right).
\end{multline}
Since this is true for all $m$, we conclude that 
\begin{multline}
\limsup_{n\to\infty} P\left(X_{n} \le 2^{-2^{(n + t\sqrt{n})/2+f(n)}}\right)\\
\le \lim_{m\to\infty} \left\{Q(t)P(X_m\le\delta)
+ P\left(X_\infty\le \frac{\delta}{2},~X_m>\delta\right)\right\}\\
= Q(t)P(X_\infty=0),
\end{multline}
holds, where we have used the almost-sure convergence 
of $X_m$ to $X_\infty$ (property 1 in Sect.~\ref{sec:rp}).  
\end{IEEEproof}

Putting Propositions~\ref{prop:direct} and \ref{prop:converse} together, 
we arrive at the following theorem.  
\begin{theorem}
\label{th:bound-Z}
For an arbitrary function $f(n)=o(\sqrt{n})$
\[
\lim_{n\to\infty} P\left(X_n\le 2^{-2^{(n+t\sqrt{n})/2+f(n)}}\right)= Q(t)P(X_\infty=0).
\]
\end{theorem}
In applying Theorem~\ref{th:bound-Z} to 
$\{Z_n\}$, it should be noted that 
$P(Z_\infty=0)=I(W)$ holds.   
Theorem~\ref{th:Pe} is proved straightforwardly on the basis of Theorem~\ref{th:bound-Z} 
via the argument at the end of Sect.~\ref{sec:prob}.  

\section{Discussion}
\label{sec:discussion}
\subsection{Extension to Construction with a Larger Matrix}
Polar codes can be constructed on the basis of a matrix larger than 
the $2\times2$ matrix $F$ in~\eqref{eq:mtxF}.  
Korada, \c{S}a\c{s}o\u{g}lu, and Urbanke~\cite{KoradaSasogluUrbanke2009} 
have provided a full characterization of whether a matrix induces 
channel polarization.  
They have shown that if an $\ell\times\ell$ matrix $G$ is polarizing, 
then given a symmetric B-MC $W$, 
\[
\lim_{n\to\infty}P\left(Z_n\le 2^{-\ell^{n\beta}}\right)=I(W)
\]
holds for any $\beta<{\tt E}(G)$, where ${\tt E}(G)$ is 
the exponent of the matrix $G$ defined in~\cite{KoradaSasogluUrbanke2009}.  
For a non-polarizing matrix, the exponent ${\tt E}(G)$ is zero.  

Our analysis can be extended to obtain a rate-dependent result 
for channel polarization using a larger matrix.  
The extension includes introduction of a sequence $\{B_i\}$ of 
i.i.d.\ random variables 
with $P(B_1=k)=1/\ell$ for $k=1,\,2,\,\ldots,\,\ell$.  
Let $\{D_1,\,D_2,\,\ldots,\,D_\ell\}$ be ``partial distances'' of 
the matrix $G$ defined in~\cite{KoradaSasogluUrbanke2009}.  
The exponent ${\tt E}(G)$ is given by the mean 
of the random variable $\log_\ell D_{B_i}$.   
Let ${\tt V}(G)$ be the variance of the random variable $\log_\ell D_{B_i}$.  
Our result in this direction is the following: 
\begin{equation}
\lim_{n\to\infty}P\left(Z_n\le2^{-\ell^{n{\tt E}(G)+t\sqrt{n{\tt V}(G)}}}\right)=Q(t)I(W).
\end{equation}
The worst case of polarizing partial distances 
is given by the case where only one of $\{D_1,\,D_2,\,\ldots,\,D_\ell\}$ 
is equal to 2 and the rest are equal to 1.  
Since ${\tt E}(G)=\frac{\log_\ell2}{\ell}$ and 
${\tt V}(G)=\left(\frac{\log_\ell2}{\ell}\right)^2(\ell-1)$ 
for the worst case,
a universal bound is obtained as 
\begin{equation}
\lim_{n\to\infty}P\left(Z_n\le2^{-\ell^{\left(n+t\sqrt{(\ell-1)n}\right)\frac{\log_\ell2}{\ell}}}\right)\ge Q(t)I(W),
\end{equation}
which can be regarded as a refinement of Theorem 8 in~\cite{KoradaSasogluUrbanke2009}.  

\subsection{Minimum Distance and ML Decoding}
We return to the original construction of polar codes on the basis of the $2\times 2$ matrix $F$.  
Polar codes are linear codes, and their generator matrices are obtained from the matrices of the form $F^{\otimes n}$ 
via removal of some rows (corresponding to ``shortening'') and reordering of 
the remaining rows.  
Hussami, Korada, and Urbanke~\cite{HussamiKoradaUrbanke2009} studied 
the class of linear codes constructed from $F^{\otimes n}$ via shortening, 
and showed using minimum distance analysis that the error probability of such codes 
is $\omega(2^{-2^{\beta n}})$ (in the standard Landau notation) for $\beta>\frac12$.
This fact means that polar codes with SC decoding achieve the best performance 
as $n\to\infty$ up to the dominant term in the double exponent of the error probability.  
In this subsection, it is shown that the minimum distance analysis does not give
the second dominant term in the double exponent of polar codes with SC decoding.
This fact implies that SC decoding is not necessarily optimal 
in the second dominant term.  

\begin{proposition}
\label{prop:md}
For any codes whose generator matrix consists of $2^nR$ distinct rows of $F^{\otimes n}$ and any fixed $t > Q^{-1}(R)$,
the error probability of ML decoding is $\omega(2^{-2^{(n + t\sqrt{n})/2}})$.
\end{proposition}
\begin{IEEEproof}
Let $\mathcal{I}\subseteq\{0,1,\dotsc,2^n-1\}$ denote the set of indices of rows of $F^{\otimes n}$ 
chosen to form the generator matrix.
The minimum distance of the codes is given by $\min_{i\in \mathcal{I}} 2^{w(i)}$,
where $w(i)$ denotes the Hamming weight of the binary expansion of $i$.
Let the minimum distance of a code be $2^{d}$.  
Since the number of rows with weight $2^i$ of the matrix $F^{\otimes n}$ 
is $\left(n\atop i\right)$, one obtains the inequality 
\begin{equation}
\sum_{i=d}^{n}\binom{n}{i} \ge 2^nR,
\end{equation}
or equivalently, 
\begin{equation}
P(S_n \ge d) \ge R,
\end{equation}
where $S_n$ is a sum of $n$ i.i.d.\ Bernoulli random variables with probability one half.
Let $d=n/2 + t\sqrt{n}/2$ for any fixed $t\in\mathbb{R}$.  
Then, 
\begin{equation}
P\left(\frac{S_n - \frac{n}2}{\frac{\sqrt{n}}2} \ge t\right) 
\ge R.
\end{equation}
From the central limit theorem, the left-hand side converges to $Q(t)$ as $n\to\infty$.
Hence, the condition $t \le Q^{-1}(R)$ is necessary for asymptotic existence of the codes 
satisfying the conditions stated in the Proposition, completing the proof.  
\end{IEEEproof}

It should be noted that Proposition~\ref{prop:md} also means that 
the minimum distance of the codes considered is asymptotically at most $2^{(n + Q^{-1}(R)\sqrt{n})/2}$.  

The prefactor of the second dominant term $\sqrt{n}$ in the double exponent 
is $Q^{-1}(R)/2$ in Proposition~\ref{prop:md}, 
which is strictly larger than the prefactor $Q^{-1}\bigl(R/I(W)\bigr)/2$ in Theorem~\ref{th:Pe} 
whenever $I(W)<1$.  
One can argue that it might be due to the channel-independent nature of the analysis 
leading to Proposition~\ref{prop:md}, which is reflected in the absence of the channel $W$ 
in the result.  
In any case, whether polar codes with SC decoding are optimal in terms of 
the double exponent up to the second dominant term is an open problem, 
and thus needs further investigation.  

\section{Conclusion}
\label{sec:conclusion}
We have derived a rate-dependent upper bound 
of the best achievable block error probability of polar codes 
with successive cancellation decoding.  
The derivation is based on that of the previous rate-independent 
results~\cite{Arikan2009,ArikanTelatar2009}, 
which discusses channel polarization 
in regimes of the law of large numbers, 
extending it to regimes of the central limit theorem.  

We would like to mention that the argument given in this paper 
can also be applied to the problem of lossy source coding 
discussed in~\cite{HussamiKoradaUrbanke2009}.

\end{document}